\documentclass[prd,aps,showpacs,amsmath,amssymb,twocolumn]{revtex4-1}
\usepackage{graphicx}% Include figure files
\usepackage{epstopdf}
\usepackage{dcolumn}% Align table columns on decimal point
\usepackage{bm}% bold math
\usepackage{color}

\begin{document}
%\LARGE
%\preprint{gr-qc/}
%\draft
\title{Energy decomposition within Einstein-Born-Infeld black holes}
\author{Jonas P. Pereira$\mbox{}^{1,2}$}
 \email{jonaspedro.pereira@icranet.org}
 \author{Jorge A. Rueda $\mbox{}^{2,3}$}
 \email{jorge.rueda@icra.it}
 %
%\author{Remo Ruffini $\mbox{}^{1,2,3}$}
 %\email{ruffini@icra.it}
%
%\author{Andrea Geralico $\mbox{}^{4,5}$}
 %\email{geralico@icra.it}
 %
\affiliation{$\mbox{}^{1}$Universit\'e de Nice Sophia Antipolis, 28 Av. de Valrose, 06103 Nice Cedex 2, France}
\affiliation{$\mbox{}^{2}$Dipartimento di Fisica and ICRA, Universit\`a di Roma ``La Sapienza'', Piazzale Aldo Moro 5, I-00185 Rome, Italy}
\affiliation{$\mbox{}^{3}$International Center for Relativistic Astrophysics Network
(ICRANet), Coordinating Center, Piazza della Repubblica 10, 65122, Pescara, Italy}
%\affiliation{$\mbox{}^{4}$Istituto per le Applicazioni del Calcolo ``M. Picone,'' CNR, I-00185 Rome, Italy}
%\affiliation{$\mbox{}^{5}$ICRA, ``Sapienza" University of Rome, I-00185 Rome, Italy}
%\affiliation{$\mbox{}^{4}$ Departamento de F\'isica, Centro de Ci\^encias Exatas e
%Tecnol\'ogicas (CCET), Universidade Estadual Vale do Acara\'u, Avenida Doutor Guarani
%317, Campus da Cidao, CEP 62.040-730, Sobral, Cear\'a, Brazil;}
%
\date{\today}

\begin{abstract}
We analyze the consequences of the recently found generalization of the Christodoulou-Ruffini black hole mass decomposition for Einstein-Born-Infeld black holes [characterized by the parameters $(Q,M,b)$, where $M=M(M_{irr},Q,b)$, $b$ scale field, $Q$ charge, $M_{irr}$ ``irreducible mass'', physically meaning the energy of a black hole when its charge is null] and their interactions.
We show in this context that their description is largely simplified and can basically be split into two families depending upon the parameter $b|Q|$. If $b|Q|\leq1/2$, then black holes could have even zero irreducible masses and they always exhibit single, non degenerated, horizons. If $b|Q|>1/2$, then an associated black hole must have a minimum irreducible mass (related to its minimum energy) and has two horizons up to a transitional irreducible mass. For larger irreducible masses, single horizon structures raise again. By assuming that black holes emit thermal uncharged scalar particles, we further show in light of the black hole mass decomposition that one satisfying $b|Q|>1/2$ takes an infinite amount of time to reach the zero temperature, settling down exactly at its minimum energy. Finally, we argue that depending on the fundamental parameter $b$, the radiation (electromagnetic and gravitational) coming from Einstein-Born-Infeld black holes could differ significantly from Einstein-Maxwell ones. Hence, it could be used to assess such a parameter.
\end{abstract}

%]
\maketitle

%\section{Introduction}
\section{Introduction}

%eliminated The first concept of a charged black hole (BH) raised after the solution of spherically symmetric
%Einstein-Maxwell equations by Reissner and Nordstr\"om \cite{1973grav.book.....M}. As it is well-known, the BH charge $Q$ modifies in
%a fundamental way the event horizon of the BH, competing with the BH mass-energy $M$. Whenever $|Q|<M$, the singularity is clothed by event horizons.

Although solving Einstein equations for a classical charged black hole (BH) (Reissner-Nordstr\"om one) is a relatively simple task \cite{1973grav.book.....M}, such an approach does not make evident the relationship between its two parameters, namely its mass ($M$) and charge ($Q$). Intuitively, this relation must exists since electromagnetic energies have their origin in charges, and it can be found in a variety of ways. An interesting, notably physical manner, was put forward by Christodoulou \cite{1970PhRvL..25.1596C} and Christodoulou and Ruffini \cite{1971PhRvD...4.3552C}, by introducing the concept of BH reversible transformations \cite{1970PhRvL..25.1596C}. Such transformations are the only ones that could bring back the BH parameters to their original values after any transformation processed by a test particle with parameters $m$ and $q$ (where $M\gg m$ and $Q\gg q$). Another known approach was due to Bardeen, Carter and Hawking \cite{1973CMaPh..31..161B}, which takes advantage of the spacetime symmetries.

It has been recently shown \cite{2014PhLB..734..396P}, in the context of spherically symmetric spacetimes, that reversible transformations are fully equivalent to the constancy of the event horizon upon such changes for any nonlinear theory of the electromagnetism $L(F)$ that leads to asymptotically flat solutions. Due to the generality of the analysis, such a constant must be $2M_{irr}$, where $M_{irr}$ is the irreducible BH mass given by the total mass-energy of the system in the uncharged case, namely when $Q=0$. Due to this fact, $M_{irr}$ must be always positive. The aforementioned equivalence allows us to exchange the problem of solving nonlinear differential equations for nonlinear theories by the problem of solving algebraic equations. This procedure works only for the cases where event horizons are present. We recall that after the seminal work of Bekenstein \cite{1973PhRvD...7.2333B}, it is known that the entropy of a black hole is equivalent to its $M_{irr}$. Nevertheless, it is more appealing to our reasoning to make use of the original concept of irreducible mass, $M_{irr}$.

The aim of this work is to elaborate on the consequences of the mass-energy decomposition for nonlinear BHs and their interactions. In order to do it,
we use the specific nonlinear theory of electromagnetism due to Born \& Infeld (BI) \cite{1934RSPSA.144..425B}. Such a theory has regained
interest due to its analogous emergence as an effective theory to String Theory \cite{1997hep.th....2087R}. It was constructed with the
purpose of remedying the singular behavior in terms of energy of a pointlike charged particle. The theory introduces a parameter $b$, identified
with the absolute upper limit of the electric field of a system when just electric aspects are present. Born and Infeld fixed this parameter by
imposing that in the Minkowski spacetime the associated electromagnetic energy coming from a point-like electron equals its rest mass (unitarian viewpoint \cite{1934RSPSA.144..425B}). Nevertheless, the dualistic viewpoint \cite{1934RSPSA.144..425B} could equally well have been assumed and the parameter $b$ should be determined by a theory relying on it, such as quantum mechanics \cite{1934RSPSA.144..425B}. Actually, the BI theory
has been applied to the description of the hydrogen atom, both non-relativistic and relativistic one \cite{2006PhRvL..96c0402C,2011PhLA..375.1391F},
and their numerical analyses show that $b$ must be much larger than the value initially proposed by BI. Notwithstanding, a definite value has not been
obtained.

Rasheed \cite{1997hep.th....2087R} has analyzed mathematically the validity of the zeroth and first laws of black hole mechanics and concluded that they do hold for any nonlinear Lagrangian of the electromagnetism. Although Rasheed concluded that the black hole mass formula for such a case does not keep the same simple functional form as for the Maxwellian Lagrangian, a further scrutiny of the consequences of this fact was not performed. Following our results in Ref.~\cite{2014PhLB..734..396P}, we instead shall analyze in this work some consequences of the black hole mass formula in the case of Einstein-Born-Infeld black holes, and their interactions. Since such a relation establishes a constraint for the parameters of the theory, physically based on conservation laws, the description is expected to be greatly simplified, as it will turn out to be exactly the case. To the best of our knowledge, this has not been done before.

%eliminated We shall here derive and analyze the properties of the Einstein-Born-Infeld BH mass-formula and some of its possible applications.
The article is organized as follows. In the next section, the mathematical approach for reversible transformations is briefly
elaborated and the mass decomposition for $L(F)$ theories in the spherically symmetric case is exhibited. In Section \ref{bi lag}, we revisit some aspects of
the Einstein-Born-Infeld black hole solution and exhibit the black hole mass decomposition for this theory. In Section \ref{clothed}, we analyze some properties of the above mentioned mass decomposition and show that when $b$ is finite, there are always intrinsic nonclassical islands of black hole solutions where each member has a single, nondegenerated horizon. Section \ref{hawking} is devoted to the study of the consequences of assuming that Einstein-Born-Infeld black holes evaporate within the framework of the mass decomposition. In Section \ref{GrR} we analyze the radiation emitted by two interacting Einstein-Born-Infeld black holes and show by means of a toy model that in principle there are alternative ways to infer the constant $b$ even from astrophysical scenarios. Section \ref{discussion} closes the paper with an analysis
of the main points raised.

Units are such that $c=G=1$ and the signature of the spacetime is $-2$.

\section{Black hole mass decomposition for any nonlinear theory}
\label{mass decomposition}
In the context of spherically symmetric solutions to general relativity minimally coupled to nonlinear Lagrangians of the electromagnetism,
it can be shown that the general solution to the metric is \cite{2013GReGr..45.1901D}
\begin{equation}
ds^2= e^{\nu(r)}dt^2-e^{-\nu(r)}dr^2-r^2(d\theta^2+\sin^2\theta d\varphi^2)\label{1},
\end{equation}
where \cite{2014PhLB..734..396P}
%\begin{widetext}
\begin{eqnarray}
e^{\nu(r)} &=& 1-\frac{2M}{r}+\frac{8\pi}{r}\int^{\infty}_{r} r'^2\,T^{0}{}_0(r')dr'\\ &=& 1 - \frac{2M}{r} + \frac{2 Q A_0 }{r}-\frac{2 \cal{N}}{r}
\label{2},
\end{eqnarray}
%\end{widetext}
%
%
%\begin{widetext}
\begin{equation}
E_r\doteq - \frac{\partial A_0}{\partial r},\;T^{\mu}{}_{\nu}= \frac{4L_FF^{\mu\beta}F_{\nu\beta}-L\delta^{\mu}{}_{\nu}}{4\pi},\; \frac{\partial \cal{N}}{\partial r}\doteq - L r^2 \label{3}.
\end{equation}
%\end{widetext}
%
%eliminated The constants of integration $M$ and $Q$ must be identified with the total mass (total energy) and charge of the system, respectively, as seen by distant observers \cite{1973grav.book.....M}.
We are assuming that the Lagrangian describing the electromagnetic interactions is $L=L(F)$, $F\doteq F^{\mu\nu}F_{\mu\nu}$, where $F_{\mu\nu}$ is the
electromagnetic field tensor \cite{1973grav.book.....M, 1975ctf..book.....L}. Besides, $L_F$ was defined as the derivative of $L(F)$ with respect to the invariant $F$ and
$T^{\mu}{}_{\nu}$ is the energy-momentum tensor of the matter fields \cite{1973grav.book.....M, 1975ctf..book.....L}, here the electromagnetic fields described by
$L(F)$. In the above expressions, $E_r$ is the radial component of the electric field and $A_0$ is its associated potential. In the expressions for $A_0$
and $\cal N$, it has been chosen a gauge where they are null at infinity. We stress that for obtaining $A_0(r)$ and ${\cal N}$$(r)$ from given $E_r(r)$ and
 $L(F)$, it is tacit one has to integrate from an arbitrary $r$ to infinity, since we are interested in black hole solutions \cite{2002CQGra..19..601B}.
The radial electric field satisfies the equation
\begin{equation}
L_F E_r r^2= - \frac{Q}{4}\;\;\mbox{or}\;\; \frac{\partial L}{\partial E_r}=\frac{Q}{r^2} \label{4}.
\end{equation}

In a spherically symmetric spacetime, infinitesimal reversible transformations are defined by
\begin{equation}
\delta M = \delta Q\,A_0(r_+)
\label{5},
\end{equation}
where $r_+$ is the outermost horizon from a given black hole theory, defined as the largest zero of Eq. (\ref{2}). For a general transformation, one has the
formal replacement ``$=\rightarrow \geq$'' in the above equation.
% eliminated Actually, Eq. (\ref{5}) encompasses the law of conservation of energy and charge in the
%case of reversible transformations for test particles interacting with a black hole. This is evidenced by the identifications: $\delta Q=q$ and $\delta
%M=E_{t.p.}$, where $E_{t.p.}$ is the conserved energy of a test particle of mass $m$ and charge $q$ in a spacetime described by
%Eqs. (\ref{1}), (\ref{2}), (\ref{3}) and (\ref{4}). Energy could always be extracted from a black hole whenever the right hand side of Eq. (\ref{5}) is
%negative. Notice that albeit we have commenced with a test particle, in the end reversible transformations just evidence the properties of
%the spacetime itself when conservation laws are taken into account. It signifies that the integration of Eq. (\ref{5}) will give us intrinsic
%information about the spacetime. For an irreversible transformation, one unavoidably has a description dependent upon the test particle
%aspects \cite{1971PhRvD...4.3552C}. We emphasize that $\delta M$ is assumed to be an exact differential, {like the internal energy of a
%thermodynamical system.}
%Hence, by means of convenient processes one could obtain the general expression for the energy decomposition for nonlinear black holes (black holes described by nonlinear Lagrangians of the electromagnetism), i.e., their equation of state. From what we have discussed previously, reversible processes are manifestly such ones.

The customary approach for obtaining the mass formula (energy decomposition) would be integrating Eq. (\ref{5}), given the outer horizon in terms of the
parameters coming from the electromagnetic
theory under interest and the spacetime. In general, it turns out to be impossible to work analytically for $L(F)$ {theories} in such a case.
Since one knows that there is a correlation between black holes and thermodynamics \cite{1973CMaPh..31..161B, 2007PhLB..652..338K}, one would suspect
that Eq. (\ref{5}) (thermodynamics) is somehow inside the equations of general relativity (or vice-versa). It can be shown easily that this is
indeed the case, provided that the outer horizon keeps constant under reversible transformations \cite{2014PhLB..734..396P}. Since it is so, it follows
that the outer horizon must be identified with its associated Schwarzschild horizon (where $Q=0$), and it will be denoted by $r_+=2M_{irr}$.

For the nonlinear theories where the electric potential $A_0$ is independent of the parameter $M$, it follows from the above reasoning and Eq. (\ref{2}) that
%
%\begin{widetext}
\begin{eqnarray}
M&=&M_{irr} + Q A_0|_{r=2M_{irr}} - {\cal N}|_{r=2M_{irr}}\nonumber \\ &=& M_{irr}+4\pi\int^{\infty}_{2M_{irr}} r'^2\,T^{0}{}_0(r')dr' \label{6}.
\end{eqnarray}
%\end{widetext}
%
The above equation is the way of decomposing the total energy in terms of intrinsic ($M_{irr}$) and extractable quantities ($M-M_{irr}$). It can  be shown with ease \cite{2014PhLB..734..396P} that it implies the so-called generalized first law of black hole mechanics for nonlinear electrodynamics \cite{1997hep.th....2087R}, thus superseding it. Notice from the above equation that
one could not associate all $M_{irr}$ (given $M$ and $T^{0}{}_{0}$) with the outer horizon. The reason for this is simple: Eq. (\ref{6}) was defined
by $e^{\nu(2M_{irr})}=0$, which encompasses also $M_{irr}$ related to the inner horizon. Nevertheless, it is uncomplicated to single out the set of $M_{irr}$ corresponding to the outer horizon. One knows that the condition that
leads to the degeneracy of the horizons is the common solution to $e^{\nu (2M_{irr})}=0$ and $d e^{\nu}/dr|_{r=2M_{irr}}=0$. These requirements
and Eq. (\ref{6}) imply that the horizons of black holes are degenerated at the critic points of $M$ as a function of $M_{irr}$. Hence, since outer horizons are larger than inner ones, it follows that the set of irreducible masses relevant in our analysis are the ones that always give $dM/dM_{irr}\geq 0$.
In the mass decomposition approach the region inside the outer horizon is not of physical relevance.

% eliminated{We highlight that just the outer horizon is of physical importance to the energy decomposition expression, Eq. (\ref{6}). This is so due to the fact that the outer horizon is the separatrix of regions where the $r$-coordinate is time-like and space-like. In the region between the inner and outer horizons, particles cannot remain still and are mandatorily impelled towards the singularity. Due to this unidirectional property, irrespective of their fate inside the inner horizon, conservation laws guarantee that just their crossing at the outer horizon will impinge changes to the black hole mass and charge in the way we described previously for external observers.}

\section{Born-Infeld Lagrangian}
\label{bi lag}
The Born--Infeld Lagrangian, $L_{BI}$, can be written as (compatible with our previous definitions)
\begin{equation}
L_{BI}=b^2\left( 1- \sqrt{1+\frac{F}{2\,b^2}}\right)\label{7},
\end{equation}
where $b$ is the fundamental parameter of the theory and counts for the maximum electric field exhibited by an electrically charged and at rest particle in flat spacetime \cite{1934RSPSA.144..425B}.
This parameter naturally defines a scale to the Born-Infeld theory.
%eliminate and thence has a non-vanishing energy--momentum tensor trace, as it could be checked from the energy momentum tensor associated with Eq. (\ref{7}) (see Ref. \cite{2010PhRvD..81f5026L} for more details).

Putting Eq. (\ref{7}) into Eqs. (\ref{2}) and (\ref{3}) and performing the integral from a given arbitrary radial coordinate $r$ up to infinity,
one gets (see for instance Ref. \cite{2002CQGra..19..601B})
\begin{equation}
e^{\nu(r)}=1-\frac{2M}{r}- \frac{2}{3}b^2y^2+\frac{2Q^2}{3\,\sqrt{|\beta|}\,r}{\cal F}\left[ x(r),\frac{1}{\sqrt{2}}\right]\label{8},
\end{equation}
where we have defined
\begin{equation}
x(r)\doteq \arccos\left( \frac{r^2-|\beta|}{r^2+|\beta|}\right),\;\;y^2\doteq \sqrt{r^4+\beta^2}-r^2\label{9},
\end{equation}
\begin{equation}
\beta^2\doteq \frac{Q^2}{b^2},\;\; {\cal F}\left[x(r),\frac{1}{\sqrt{2}}\right]=2\int^{\infty}_{\frac{r}{\sqrt{|\beta|}}}\frac{du}{\sqrt{1+u^4}}\label{10},
\end{equation}
where ${\cal F}[x(r),1/\sqrt{2}]$ is the elliptic function of first kind \cite{2007tisp.book.....G}.

The modulus of the radial electric field and its scalar potential in this case, as given by the first term of Eq. (\ref{3}) and Eq. (\ref{4}), are
\begin{equation}
E_r(r)=\frac{Q}{\sqrt{r^4+\beta^2}},\;\; A_0(r)=\frac{Q}{2\,\sqrt{|\beta|}}{\cal F}\left[x(r),\frac{1}{\sqrt{2}}\right]\label{11}.
\end{equation}
As it is clear from Eq. (\ref{11}), the electric field of a pointlike charged particle is always finite, as well as its associated scalar potential
and they are positive monotonically decreasing functions of the radial coordinate. Hence, from Eq. (\ref{5}), it implies that the necessary and
sufficient condition for extracting energy from a Einstein-Born-Infeld black hole is to use test particles with an opposite charge to the hole.

\section{Analysis of the Einstein-Born-Infeld mass formula}
\label{clothed}

The metric given by Eqs. (\ref{8}), (\ref{9}) and (\ref{10}) has been studied in detail in Ref. \cite{2002CQGra..19..601B}. It has been pointed out there
that the dimensionless
quantities
$\tilde{M}\doteq bM$, $\alpha\doteq Q/M$ and $u\doteq r/M$ are convenient to scrutinize the properties of such a metric. Nevertheless, apparently some
interesting
properties of Eq. (\ref{8}) have not been stressed.
Under the above definitions, Eq. (\ref{8}) may be written as
%
%\begin{widetext}
\begin{eqnarray}
e^{\nu(u)}&=& 1-\frac{2}{u}+\frac{2}{3}\tilde{M}^2u^2\left( 1-\sqrt{1+\frac{\alpha^2}{\tilde{M}^2u^4}}\right)  \nonumber\\ &+& \frac{2\alpha^2}{3u}\sqrt{\frac{\tilde{M}}{|\alpha|}}{\cal F}\left[ \arccos\left({\frac{\tilde{M}u^2-|\alpha|}{\tilde{M}u^2+|\alpha|}}\right),\frac{1}{\sqrt{2}}\right]\label{11a}.
\end{eqnarray}
%\end{widetext}
%
The horizons are obtained as the zeros of the above equation.
%In order to investigate them, the derivative of Eq. (\ref{11a}) is also important.
As a result, one can verify that Eq. (\ref{11a}) has no minimum, and hence it is a monotonic function iff
\begin{equation}
b< \frac{9M^2}{|Q|^3\,{\cal F}^2\left[\pi,\,\frac{1}{\sqrt{2}}\right]}\approx \frac{0.654 M^2}{|Q|^3}\label{12},
\end{equation}
{which can also be cast as
\begin{equation}
M>M_0,\;\;M_0\doteq \frac{\sqrt{b|Q|^3}}{3} {\cal F}\left[\pi,\frac{1}{\sqrt{2}}\right] \label{born_mass}.
\end{equation}
}
As the limit of $u$ going to zero in Eq. (\ref{11a}) shows us, Eq. (\ref{12}) also guarantees that the associated spacetime will always exhibit just one horizon (not degenerated). The above inequality has no classical counterpart, since it can be formally obtained by taking the limit of $b$ going to infinity. Equation (\ref{12}) sets a fundamental inequality concerning the parameters $Q$, $b$ and $M$.
Whenever it is not verified, it does automatically imply the existence of a minimum.
%eliminated Nevertheless, as it is easy to be realized, it does not imply the existence of horizons. Hence another more restrictive condition must exist assuring it.
A simple analysis shows us that such a requirement can be cast as
\begin{equation}
u_+\leq \frac{\sqrt{4\tilde{M}^2\alpha^2-1}}{2\tilde{M}},\;\;\;\frac{d}{du}(e^{\nu})|_{u=u_+}=0\label{12a},
\end{equation}
{which is just the consequence of imposing that $e^{\nu(u_+)}\leq 0$, $u_+$ being the critical point of $e^{\nu}$, thus guaranteeing the existence of an outer horizon.}
Just as a reference, in the limit when $\tilde{M}$ goes to infinity, the above condition reduces to $|\alpha|\leq 1$, as it is well-known from the Reissner-Nordstr\"om solution for assuring the existence of horizons. As the above inequality suggests, the term $(4b^2Q^2-1)$ plays a fundamental role into the horizon description. We shall see that this is also the case in the approach related to the energy decomposition.
%eliminated, as one would expect from our previous reasoning.
Specialized to the Born-Infeld Lagrangian, Eq. (\ref{7}), the total mass [see Eq. (\ref{6})] of an Einstein-Born-Infeld black hole can be decomposed as
%
%\begin{widetext}
\begin{eqnarray}
M &=& M_{irr}-\frac{8}{3}b^2M_{irr}^3\left(\sqrt{1+\frac{\beta^2}{16M_{irr}^4}}-1\right) \nonumber\\ &+& \frac{\sqrt{b |Q|^3}}{3}{\cal F}\left[\arccos\left(\frac{4M_{irr}^2-|\beta|}{4M_{irr}^2+|\beta|}\right),\frac{1}{\sqrt{2}}\right]\label{13}.
\end{eqnarray}
%\end{widetext}
%
%eliminated We recall that the validity of the above energy expression automatically ensures the existence of an outer horizon. One should though keep in mind that just a subset of all possible $M_{irr}$ are related to the aforementioned horizon and they are the values that always lead to $dM/dM_{irr}\geq 0$. The nonexistence of a horizon here means the impossibility of finding an irreducible mass for a given energy. In the next section we shall take a closer look to Eq. (\ref{13}).

%\textcolor{red}{This paragraph was originally in the section of Hawking temperature.}
%eliminated It is well-known that any nonlinear theory of the electromagnetism has a nonlinear medium associated with it \cite{1934RSPSA.144..425B} in the scope of the Maxwellian electromagnetism (the converse is not necessarily true! See Ref. \cite{2014PhRvA..89d3822D} for further details). Thence, one expects the extractable energy of a charged
%nonlinear black hole described by Born-Infeld theory to be smaller than its classical counterpart. This is corroborated by the case the fields are
%much smaller than the scale field $b$ \cite{2014PhLB..734..396P}. It can be verified this is indeed always the case for those black holes that generalize the
%Reissner-Nordstr\"om ones.
%One should simply take convenient limits on the ratio of the Born-Infeld and Maxwell total energies and check that they tend either to zero or to a value smaller than the unity.

\begin{figure}[!htbp]
\centering
\includegraphics[width=\hsize,clip]{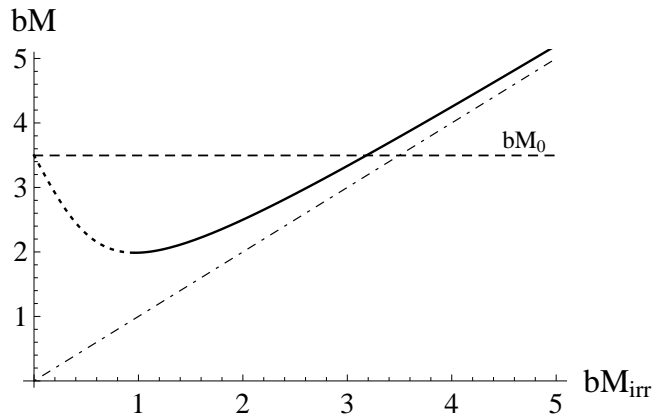}
\caption{{\small\sf Mass formula (thick plus dotted curves), Eq. (\ref{13}), when the parameter $b|Q|$ satisfies Eq. (\ref{14}),
chosen here as $2$. The dashed curve represents $bM_{0}$, as given by Eq. (\ref{born_mass}). The dot-dashed curve is the asymptote to
$M$, $M_{irr}$. Besides, $bM$ exhibits a minimum at the critical point $M_{irr}^{c}b\approx 0.97 $ (where the horizons become degenerated)
and for $M_{irr}^c b\leq M_{irr}b< M_{irr}^tb\approx 3.18$, we have the range of irreducible masses that generalize Reissner-Nordstr\"om black holes.
For $M_{irr}\geq M_{irr}^t$, there is a sole horizon (not degenerated), whose radial coordinate inside of it is always space-like. The irreducible masses associated with the
outer horizon are $M_{irr}\geq M_{irr}^c$. The dotted curve is related to the inner horizon solutions (for given configurations) and
are not relevant  to the analyses {concerning the black hole mass decomposition}.}}
\label{fig1}
\end{figure}

%
From now on we shall assume that Eq. (\ref{13}) is a valid decomposition to the total energy of a Einstein-Born-Infeld black hole.
A simple analysis tells us that whenever
\begin{equation}
2b|Q|>1  \label{14}
\end{equation}
is valid for the parameter $Q$, given $b$, Eq. (\ref{13}) does have a minimum with respect to $M_{irr}$, associated with the critical irreducible mass
\begin{equation}
M_{irr}^{c}\equiv M_{irr}^{min}= \frac{\sqrt{4 b^2 Q^2-1}}{4 b} \label{mirredc}.
\end{equation}
Note that $M_{irr}^c$ {is always related to the case where the horizons are degenerated (extreme black holes), as we have pointed out in section \ref{mass decomposition}, and it} is always smaller than its classical counterpart, $|Q|/2$ {(where $M=|Q|$).} From our previous discussions, the relevant irreducible masses to the analysis for reversible transformations for black holes are $M_{irr}\geq M_{irr}^c$.
Substituting the above critical irreducible mass into Eq. (\ref{13}), one has that such a minimum total energy is
%
%\begin{widetext}
\begin{equation}
M_{min}=\frac{\sqrt{4b^2Q^2-1}}{6b}+\frac{\sqrt{b|Q|^3}}{3}{\cal F}\left[x\left(\frac{\sqrt{4b^2Q^2-1}}{2b}\right) ,\frac{1}{\sqrt{2}}\right]\label{15},
\end{equation}
%\end{widetext}
%
which is naturally positive and it can be verified to be smaller than {$M_0$ defined by Eq. (\ref{born_mass}).}
{For the case $2b|Q|>1$, one can check that an immediate solution to  $M=M_0$ is} $M_{irr}=0$ (not of relevance for us for the present case). {There also is a nontrivial solution that cannot be expressed analytically in general, that we shall denote by $M_{irr}^t$.} This solution is very important {since} it {will} delimit the transition from space-like singularities to time-like ones with respect to the radial coordinate. This signifies that the range of irreducible masses that generalizes Reissner-Nordstr\"{o}m black holes (with two horizons) is $M_{irr}^{min}\leq M_{irr}<M_{irr}^t$. An arbitrary black hole with $M_{irr}\geq M_{irr}^t$ shall present a sole horizon and hence when test particles have crossed it, their fate is unavoidably its associated singularity. Note that Reissner-Nordstr\"om black holes are such that $M_{irr}^t\rightarrow \infty$ and the existence of $M_{rr}^t$ for Einstein-Born-Infeld black holes is only due to the finiteness of $b$. Figure \ref{fig1} exemplifies the analysis from the previous sentences for
a selected value of the parameter $b|Q|$ {for the case $2b|Q|>1$.}

\begin{figure}[!htbp]
\centering
\includegraphics[width=\hsize,clip]{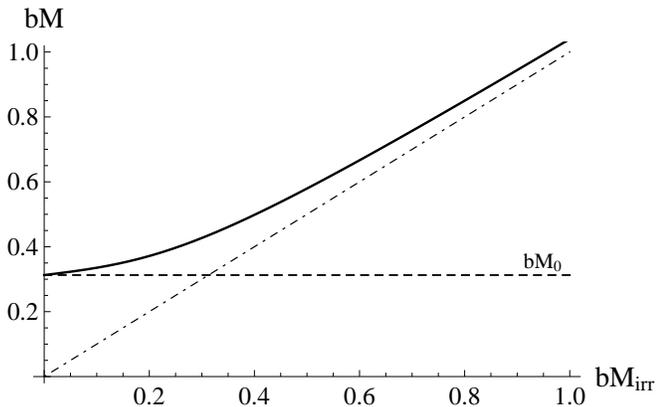}
\caption{{\small\sf Mass decomposition when the parameter $b|Q|$ does not satisfy Eq. (\ref{14}) and is chosen to be $0.4$.
The curves have the same meaning as the ones in Fig. \ref{fig1}. From the solid curve we see that $M$ is a monotonic function and always larger than $M_0$. This means
that such a case characterizes a scenario where there is always a sole event horizon and there is no classical
analogue to it.}}
\label{fig2}
\end{figure}

\begin{figure}[!htbp]
\centering
\includegraphics[width=\hsize,clip]{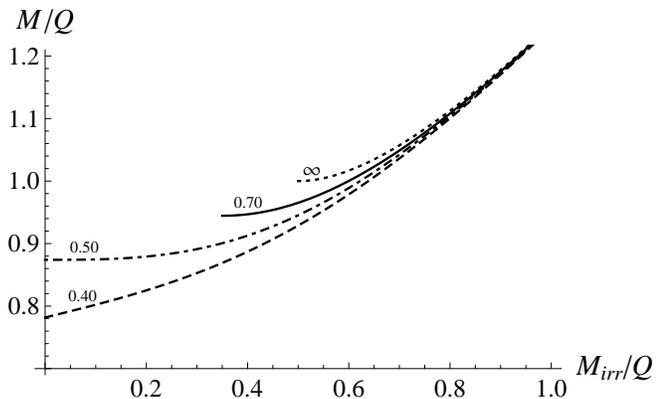}
\caption{{\small\sf Mass formula for selected values of the parameter $b|Q|$ (numbers on the curves) that encompasses all physically distinct classes of black
holes for the Born-Infeld Lagrangian. The dotted curve represents the mass formula for the Maxwell Lagrangian.
The dot-dashed curve demarcates the transition from two horizon solutions (as given by the thick curve) to a single one (as given by the dashed curve),
where its associated inner horizon is null. The branches related to the inner horizons were removed. }}
\label{mqcc}
\end{figure}

We consider now the case where Eq. (\ref{14}) is violated. In this case, $M$, as given by Eq. (\ref{13}), is a monotonic function of
$M_{irr}$. Since it is given by Eq. (\ref{born_mass}) when $M_{irr}=0$ and it is monotonic, we conclude that Eq. (\ref{born_mass}) is
always satisfied and therefore the associated singularity is unavoidable for test particles.
%eliminated Hence, whenever Eq. (\ref{14}) is not satisfied, one is led unavoidably to a singularity with just one horizon, for any value the associated irreducible mass of the system may have. This means that once test particles enter the horizon of the aforesaid BHs, the singularity is unavoidable for them.
Just for completeness, Fig. \ref{fig2}
compactifies {the above mentioned properties} for a selected value of the parameter $b|Q|$ {such that $2b|Q|\leq 1$}. Besides, in Fig \ref{mqcc} we depicted all the different classes associated with the parameter $b|Q|$,
assuming in all cases it is fixed.

An important general remark is here in order, specially for astrophysical analyses. Assume that $b|Q|=C_1$ and $M/|Q|=C_2$, where $C_1$ and $C_2$ are given constants. This means that $Mb=C_1 C_2$, is also known.
%eliminated Nevertheless, from our previous analyses with the generalized Christodoulou-Ruffini mass formula for the Born-Infeld Lagrangian, it is
%simple and also intuitive to see that $bM\geq (bM)_{min}\geq 0$.
Assuming that $0\leq C_1<\infty$ and from the fact that $bM\geq (bM)_{min}\geq 0$, we firstly conclude that $C_2$ cannot be any, but $C_2\geq (bM)_{min}/C_1$.
This means that $|\alpha|\doteq |Q|/M\leq C_1/(bM)_{min}$ and this is the condition that guarantees the presence of an outer horizon in an Einstein-Born-Infeld
black hole. In the classical case for instance, where $(bM)_{min}= b|Q|$ [see Eq. (\ref{15}) in the limit $b\rightarrow \infty$], the previous inequality
means $|\alpha|\leq 1$, as it is already known.
%eliminated Let us consider that we have chosen a convenient value for $C_2$. Thus, the mass formula always allows one to know $bM_{irr}$.
Finally, after one chooses arbitrarily another parameter, be $M$ or $|Q|$ or $b$, all the remaining ones are automatically fixed,
which could be assessed by the aforesaid choice.
%eliminated For example, if $M$ is chosen, which should usually be the case for applying theoretical analyses to astrophysical systems since this parameter can be obtained by other means, one could see the consequences e.g. on the radiation coming from the coalescence of charged black holes for several values of charge and $b$ (by changing $C_1$ and $C_2$ in accord with what has just been explained above) and then assess their relevance.
It is not complicated to see that when $M_{irr}/|Q|$ is given instead of $M/|Q|$, a similar reasoning as the above one also ensues.

\section{Hawking radiation from Einstein-Born-Infeld black holes}
\label{hawking}
\begin{figure}[!htbp]
\centering
\includegraphics[width=\hsize,clip]{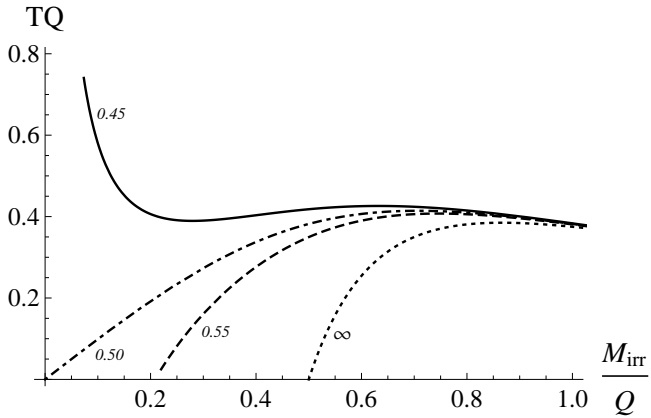}
\caption{{\small\sf Einstein-Born-Infeld black hole temperature as a function of the irreducible mass for selected values of the parameter $b|Q|$.
The temperature goes to infinity as the irreducible mass tends to zero whenever $2b|Q|\leq 1$ (thick curve). Whenever $2b|Q|>1$ (dashed
curve), it decreases with the decrease of the irreducible mass (keeping the charge constant), always being null for a finite value of the latter. The temperature experiences a transitional behavior for $2b|Q|=1$ (dot-dashed curve), being null just when the irreducible mass of the system is so {[see Eq. (\ref{temp})], albeit it cannot be seen directly from this plot}.
Finally, $b|Q|\rightarrow \infty$ (dotted curve) corresponds to the Reissner-Nordstr\"om case. }}
\label{tempqconst}
\end{figure}

{Subsequent to the work of Hawking on the semiclassical quantization of scalar fields in some curved spacetimes \cite{1975CMaPh..43..199H}, it is widely accepted that black holes radiate thermally, although this view has still some criticisms \cite{1995PhLA..209...13B, 2006PhLA..354..249B}. Motivated by the first law of black hole thermodynamics, which is a direct consequence of the mass decomposition expression given by  Eq. (\ref{6}) \cite{2014PhLB..734..396P}, and the results from the aforesaid semiclassical quantization,} we shall now study the consequences of conjecturing that clothed black holes should behave like blackbodies to observers at infinity (no back-reaction effects are considered here), radiating at temperatures proportional to their surface gravity \cite{1975CMaPh..43..199H}. In the spherically symmetric case, such a
quantity is proportional to $d e^{\nu}/dr|_{r=r_+}$ \cite{2007PhLB..652..338K, 2004reto.book.....P}. From Eq. (\ref{11a}) and preceding definitions, one has
\begin{equation}
T\propto \frac{1 + 8 b^2 M_{irr}^2 -2 b \sqrt{16 b^2 M_{irr}^4+Q^2}}{M_{irr}}\label{temp}.
\end{equation}
We notice some particularities of the insertion of the parameter $b$ into the description of the electromagnetic fields. As in the classical case,
$b\rightarrow \infty$, it is possible to attain $T=0$, but now as far as
\begin{equation}
M_{irr}^{(T=0)}= \frac{\sqrt{4 b^2 Q^2-1}}{4 b}\label{bvalue}.
\end{equation}
Notice that $M_{irr}^c = M_{irr}^{(T=0)}$. This is not surprising, since from our previous comments, the condition for null temperature of a black hole with charge $Q$ occurs exactly at the critical points of the energy with respect to its irreducible mass. When Eq. (\ref{14}) holds, one sees that the temperatures of the associated clothed black holes must decrease with the decrease of their irreducible masses until they eventually reach zero, for $M_{irr}=M_{irr}^{(T=0)}$. This would mean that black holes where Eq. (\ref{14}) is valid should radiate off finite amounts of energy, namely $M(M_{irr})-M\left(M_{irr}^{(T=0)}\right)$. Besides, from the analyses of the energy decomposition, black holes could never have negative temperatures.
For the case Eq. (\ref{14}) does not hold, it is impossible to have $T=0$ and the temperature increases with the decrease of the irreducible mass. Figure \ref{tempqconst}
compactifies the dependence of the temperature upon the irreducible mass for selected values of $b|Q|$.

We elaborate now on the temperature evolution of evaporating black bodies. For an arbitrary black hole case where $2b|Q|> 1$, as we know,
the temperature decreases as the irreducible mass of the system does so [see Fig. \ref{tempqconst}]. Hence, it would allow us to conceive a situation where just the
emission of uncharged scalar particles are present. For this simplified case, {the charge of a hole would remain constant}. Given that the black
holes would behave like blackbodies for observers located at infinity (where there is a meaning to talk about the total energy of a black hole),
their energy loss could be estimated by Stefan's law \cite{1984ucp..book.....W}
\begin{equation}
\frac{dM}{d\lambda}= - M_{irr}^2 T^4\label{Steplaw},
\end{equation}
where $\lambda$ is proportional to the observer's time receiving the radiation. For the emission of uncharged scalar particles, the above equation and Eq. (\ref{13})
imply that
\begin{equation}
\frac{d\tilde{M}_{irr}}{d\lambda}\propto -\frac{\left(1 + 8 \tilde{M}_{irr}^2 -2\sqrt{16 \tilde{M}_{irr}^4+\tilde{Q}^2}\right)^3}{ \tilde{M}_{irr}^2}\label{evapeq}.
\end{equation}
In the above equation, for an arbitrary quantity $A$, $\tilde{A}\doteq bA$. We show now that for this case the temperature never reaches
the absolute zero. Since the irreducible mass can decrease until $M_{irr}^{min}$, after a convenient transient time interval,
the right hand side of Eq. (\ref{evapeq}) can always be expanded about $M_{irr}^{min}$, leading to
\begin{equation}
\frac{d\tilde{M}_{irr}}{d\bar{\lambda}}= - \left(\tilde{M_{irr}}-\frac{1}{4} \sqrt{4 \tilde{Q}^2-1}\right)^3\label{evapeqnzero},
\end{equation}
where $\bar\lambda$ is proportional to $\lambda$ and other terms that are constants and not important to our analysis. The above equation has an analytic solution and when the limit of $\bar \lambda$ going to infinity is taken, one obtains $\tilde{M}_{irr}(\infty)=\tilde{M}_{irr}^{min}$. This means an associated black hole never reaches the absolute zero and tends asymptotically to have just one horizon.
Our analyses in light of the energy decomposition gives the same known mathematical results for the thermodynamics for Reissner-Nordstr\"om black holes \cite{1984ucp..book.....W, 1986PhRvL..57..397I},
but in a simpler way. 
%eliminated The previous description naturally generalizes the analyses for Reissner-Nordstr\"om black holes \cite{1984ucp..book.....W, 1986PhRvL..57..397I}.

For an arbitrary black hole satisfying $2b|Q|\leq 1$, it seems that a juncture shall arrive where its thermal energy will be sufficient to create pairs that could even neutralize the hole. This would befall since in this case the thermal energy of a black hole would augment with the diminution of its irreducible mass [see Fig \ref{tempqconst}]. Hence its description would be much more elaborated than the former one. Black holes with $2b|Q|\leq 1$ are expected to evaporate after finite amounts of time, as corroborated by numerical analyses from Eq. (\ref{evapeq}). We shall not pursue further into these issues in this work.

%eliminatedWe further point out that from Eq. (\ref{15}) for $1/(2|Q|)\leq b\leq \infty$, one has that
%
%\begin{equation}
%\frac{1}{3\sqrt{2}}{\cal F}\left[\pi, \frac{1}{\sqrt{2}} \right](\approx 0.874) \leq\frac{M_{min}}{|Q|}\leq 1 \label{mminrange}.
%\end{equation}
%
%while that for $0<b<1/(2|Q|)$, we have
%
%\begin{equation}
%0<\frac{M_{min}}{|Q|}<\frac{1}{3\sqrt{2}}F\left[\pi, \frac{1}{\sqrt{2}} \right]\approx 0.874 \label{mminrange2},
%\end{equation}
%
%This means that the electron, with
%$|Q|\doteq |e|\approx 10^{-34}\,cm$ and $M\approx 10^{-55}\, cm$, when seen as a fundamental clothed black hole, would {necessarily} imply $b< 1/(2|e|)\approx 10^{33}cm^{-1}$. As it can be verified, in order to obtain the ratio $M/|Q|$ compatible with the electron, one should assume $b<b_0$, $b_0\approx 10^{-9}cm^{-1}$ the value obtained by Born and
%Infeld under the unitarian viewpoint. This is disagreement with the results from the hydrogen atom being described by the Born-Infeld Lagrangian, {where the description of its energy spectrum requires a value for $b$ larger than $b_0$ \cite{2006PhRvL..96c0402C, 2011PhLA..375.1391F}}. Hence, the electron should still be seen as a naked singularity and it would be meaningless to try to ascribe a temperature
%to it.

\section{Energy loss of interacting Einstein-Born-Infeld black holes}
\label{GrR}
In this section we shall make use of the energy decomposition given by Eq. (\ref{13}) to find the imprint the parameter $b$ has on the energy radiated off by two interacting Einstein-Born-Infeld black holes. For accomplishing such a goal, we shall also utilize the second law of black hole mechanics \cite{1973grav.book.....M, 1973CMaPh..31..161B}. Such a theorem implies that the area of the resultant black hole can never be smaller than the sum of the areas of the initially (far away) interacting black holes \cite{1973grav.book.....M, 1973CMaPh..31..161B}. For simplifying the reasoning, we will assume that all the black holes involved are spherically symmetric Einstein-Born-Infeld ones. This problem can easily be solved for Einstein-Maxwell black holes (Einstein theory minimally coupled to the Maxwell Lagrangian), because their outer horizons are analytical. For nonlinear black holes, in general just numerical solutions are possible. In the mass decomposition approach, it is possible to carry out the analytical investigations further. The key for this is that whenever the mass formula is taken into account, the outer horizon must be always proportional to its associated irreducible mass for any theory.

Assume that the two initially interacting black holes have irreducible masses $M_{i1}$ and  $M_{i2}$, respectively, giving rise to another (final) one of the same kind with irreducible mass $M_{if}$.
Concerning its final charge, if one assumes that just radiation is allowed to leave the system (carried away by neutral particles), it must be the sum of the charges of the two initial black holes \cite{1973grav.book.....M}. Since the irreducible masses
are proportional to the horizon areas, Hawking's theorem (or the second law of black hole mechanics) implies that
\begin{equation}
M_{if}^2\geq M_{i1}^2+M_{i2}^2\label{16},
\end{equation}
%
%eliminated A comment here is in order. Since we assumed that the initial black holes are very far away, the total energy of the system is ideally $E_t=M_1+M_2$, where $M_a$, $a=1,2,f$, is the total energy of the a{\it th} black hole.

Invoking the first law of black hole mechanics for an isolated system
\cite{1973grav.book.....M}, the final energy of the two interacting black holes, $M_f$, can never be larger than $M_1+M_2$. The difference in the energy
balance is due to the emission of radiation (here gravitational and electromagnetic), hence, $W_{rad}=M_1+M_2-M_f\geq 0$. By the cognizance of the minimum final energy
of the system, it is even possible to obtain its maximum energy radiated off, a point we shall not pursue here.
%In order not to violate the first law, just a subset of the solutions given by Eq. (\ref{16}) is physically meaningful. Naturally this is the case once we are working with a restrictive condition for the resulting black hole and miscellaneous final configurations could raise.

For fixing ideas, let us analyze first the classical case, namely two Reissner-Nordstr\"om black holes interacting in a way to lead to another Reissner-Nordstr\"om black hole. We know that the total energy of each black hole
can be written as \cite{1971PhRvD...4.3552C}
\begin{equation}
M_a=M_{ia}+\frac{Q^2_a}{4M_{ia}}\label{17},
\end{equation}
where we have defined $Q_a$ as the charge of the a{\it th} black hole.
It is easy to see that just $M_{if}^{-}\leq M_{if}\leq M_{if}^{+}$, with
%
%\begin{widetext}
\begin{equation}
M_{if}^{\pm}=\frac{M_{1}+M_{2} \pm \sqrt{(M_{1}+M_{2})^2-(Q_1+Q_2)^2}}{2} \label{18}
\end{equation}
%\end{widetext}
%
is in agreement with the above mentioned positivity of $W_{rad}$. Naturally, choices for $M_{if}$ must satisfy simultaneously Eqs. (\ref{16}) and (\ref{18}).
When nonlinear theories are present, it is clear that in general the above range of final irreducible masses will not agree with the classical (Einstein-Maxwell black holes) case. It means that many possible classical situations will not exist in the nonlinear case and vice-versa even in the simple case of symmetry conserved binary interactions. This could possibly lead to significant deviations for the amounts of radiation emitted by some systems when they are treated classically or not.

In the Einstein-Born-Infeld theory, the physical interval for $M_{if}$ cannot be determined (numerically) unless the fundamental parameter $b$ is
given. What is known \cite{2006PhRvL..96c0402C} is that $b>b_0\approx 10^{-9}{\it cm^{-1}}$, where $b_0$ is the value for the scale field determined by Born and Infeld using the unitarian viewpoint \cite{1934RSPSA.144..425B}.

Let us take a closer look at the Einstein-Born-Infeld black holes when compared to their classical counterparts. Assume just for simplicity that
$M_{i1}=M_{i2}$ and $Q_1=Q_2\equiv Q>0$. For this choice, Eq. (\ref{18}) gives us $-\sqrt{1/\alpha^2-1}\leq M_{if}/Q-1/\alpha \leq \sqrt{1/\alpha^2-1}$, where $\alpha$ is here defined as the charge-to-mass ratio of the initially interacting black holes. Let us choose, just for simpleness, $M_{if}/Q=1/\alpha$. From the Einstein-Maxwell case, one can check easily that for the above analysis $W_{rad(clas)}/Q=(1-\alpha^2)/\alpha$. For the above choice of parameters, one can show that Eq. (\ref{16}) is just satisfied if $\alpha\geq \sqrt{2(\sqrt{2}-1)}\approx 0.91$. Such cases are of theoretical
interest {since} they would evidence the departures of the Born-Infeld theory from the Maxwell theory. For investigating smaller values of $\alpha$, one should select different final irreducible masses for the black holes.

\begin{figure}[!htbp]
\centering
\includegraphics[width=\hsize,clip]{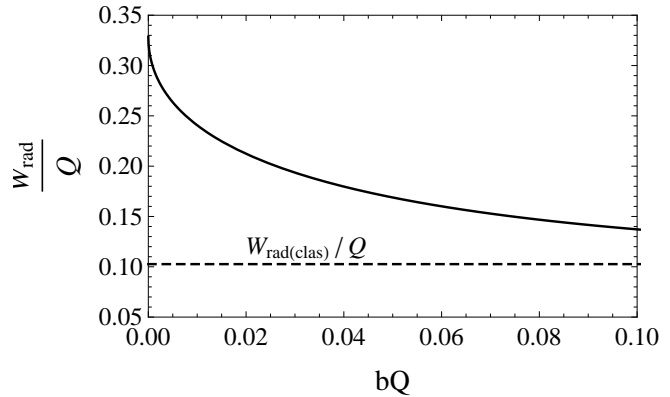}
\caption{{\small\sf Total radiation (gravitational plus electromagnetic) $W_{rad}/Q$ released in the process of coalescence of two identical Einstein-Born-Infeld black holes with $\alpha=0.95$ under the assumption it leads to another one of the same type with the same parameters as their classical counterparts. The thick curve represents such a case. The dashed curve stands for the radiation encountered in the Einstein-Maxwell theory, $W_{rad (clas)}/Q$. The associated radiation tends to its classical counterpart when $bQ$ goes to infinity. The energy released in the case of nonlinear black hole interaction is always larger than the one coming from its classical counterpart, for a given charge $Q$.
}}
\label{fig3}
\end{figure}

Figure \ref{fig3} compactifies the possibilities for the above chosen $M_{if}$ for $\alpha=0.95$, due to miscellaneous values of $bQ$. One sees in this case that
nonlinear and linear black holes may radiate off very different amounts of energy. Besides, the energy released for interacting Born-Infeld
black holes is always larger than its Maxwellian counterpart. Notice finally that $Q=\alpha M$,
$M$ being the mass of any of the black holes when they are far
apart, which would also allow one to compare the energies radiated off by the black holes during their process of interaction with the total
initial energy of the system.

Some simple estimates can be done here assessing astrophysical scenarios where Fig. \ref{fig3} could be of relevance. As we stressed before, from the hydrogen atom one knows that $b\gg b_0\simeq 10^{-9}cm^{-1}\simeq 10^{15}$ e.s.u. We also commented at the end of section \ref{clothed} that with fixed $M_{irr}/|Q|$ or $M/|Q|$ and $b|Q|$, one still has freedom to choose arbitrarily another parameter, such as $M$, even having already taken into account the mass formula. Let us choose, as it is reasonable under the point of view of black hole interactions coming from neutron stars, $M\simeq M_{\bigodot}\simeq 1.48 \times 10^{5}$cm, where $M_{\bigodot}$ is the mass of the sun. Let us focus our attention at a given value of $b|Q|$ such that the associated radiated energy may differ considerably from its classical counterpart. As a simple inspection in Fig. \ref{fig3} reveals, one could take as a good example for so $b|Q|= 0.1$, where the energy radiated off by Born-Infeld black holes is around $30\%$ more than Maxwellian ones with identical parameters. Besides we recall that we have already chosen $\alpha= 0.95$ for plotting Fig. \ref{fig3}. From this case, we have $bM_0\simeq 4\times 10^{-2}$ [see Eq. (\ref{born_mass})], which shows that $|\alpha|=0.95$ is a perfectly good candidate for the case $2b|Q|\leq 1$, the one we are interested here. For this case we know that $Q=M/C_2= 1.4\times 10^{5}$cm$=1.6\times 10^{20}$C and finally $b= C_1C_2/M= 7.1\times 10^{-7}$$cm^{-1}$, which is about a thousand times larger than $b_0$ and hence in agreement with the bound given by the hydrogen atom, the only remaining physical constraint. Therefore, the above example suggests that the radiation coming from coalescing astrophysical black holes could be a good tool to access and discriminate their electrodynamical properties.

\section{Discussion}
\label{discussion}

Foremost, it is clear {that} the approach of analyzing a given black hole solution just from its metric and the one from its metric and energy decomposition expression must be consistent since both approaches use intrinsic properties of the spacetime. Nevertheless, the latter approach is much more restrictive than the former one.
%eliminated This is in full analogy with all possible equations of state for thermodynamic systems and the ones that satisfy the first law.
It must be stressed that the energy decomposition (black hole thermodynamics) is mandatory for the proper description of any (clothed) black hole phenomenon, since it is in accord with conservation laws. Such a constraint equation (energy decomposition) in turn automatically evidences the physically relevant cases in black hole physics, hence leading to a pellucid description of them.

The energy decomposition analysis within Einstein-Born-Infeld black holes leads us to their split into two fundamental families of black holes. Whenever $2b|Q|\leq 1$, independent of their irreducible masses, one is led to an associated black hole whose singularity cannot be forestalled after test particles cross its sole, non degenerated horizon.
%eliminated Naturally this befalls {since} due to Eq. (\ref{13}) all the associated masses for the black holes must always be larger than $M_0$, defined by Eq. (\ref{born_mass}), which automatically warrants the validity of Eq. (\ref{born_mass}).
Besides, the previous inequality naturally leads to an absolute upper limit to the charge of approximately $10^{8}$cm $\approx 10^{3}M_{\bigodot}$ $\simeq 10^{23}$C, given that  $b> 10^{-9}cm^{-1}$ \cite{2006PhRvL..96c0402C}. Finally, we notice that for this class of black holes, the extractable energy could be up to one hundred percent, since black holes with $2b|Q|\leq 1$ could even have $M_{irr}=0$.
%eliminated For this case, the available energy is $M_{0}$, as given by Eq. (\ref{born_mass}) and after it has been extracted, nothing is left.
We stress that the previous conclusions are strictly nonclassical consequences of the finiteness of $b$.

The second family of black holes is defined by those satisfying $2b|Q|>1$, where $M_{irr}\geq M_{irr}^{min}$ [see Eqs. (\ref{mirredc}) and (\ref{15})] for each black hole associated. It constitutes the family that generalizes Einstein-Reissner-Nordstr\"om black holes for irreducible masses smaller than transitional values, the nontrivial solutions of $M=M_{0}$, and larger than $M_{irr}^{min}$ (related to their minimum energies), whose associated energies (masses) are always smaller than $M_0$. Above such  transitional irreducible masses, again due to the finiteness of $b$, nonclassical black holes with single horizons also rise, all of them having masses larger than $M_0$. The total amount of energy that could be extracted [$M-M_{irr}$, see Eq. (\ref{13})] in this case is always inferior to half of the total energy of the hole (as it occurs for Reissner-Nordstr\"om black holes, see \cite{1971PhRvD...4.3552C}), here due to the self-interactions present.

%eliminated Our analyses showed that the minimum irreducible masses of black holes satisfying $2b|Q|>1$ should decrease concerning their classical counterparts.
Black holes satisfying  $2b|Q|>1$ should radiate off (suppose by emitting uncharged scalar particles) until their temperatures reach $T=0$, taking for doing so an infinite amount of time, settling down exactly at their lowest energy state,
as one would intuitively expect and here as a direct consequence of the mass formula.
%eliminated This is in full agreement with the unattainability of null temperatures for thermodynamic systems and their tendency to be in their lowest energy states that are in congruence with the processes under consideration, crystalline results when the mass formula is taken into account.
Further energy could be extracted from them (obviously by means of other processes rather than the emission of uncharged scalar particles) even when $T=0$, {since} they still have an ergosphere. For the case $2b|Q|\leq 1$, it is impossible to have $T=0$ and they are expected to keep radiating, with a much more complex dynamics, until their total evaporation likely after a finite amount of time as measured by the observer who receives the radiation. Whenever charged scalar fields are taken into account, the phenomenon of superradiance could also take place, rendering their dynamics even more cumbersome. Superradiance is of interest for charged nonlinear black holes, since it is another energy extraction mechanism for them and would couple to the nonlinearities of the electromagnetic field. We let more precise analyses of this case to be done elsewhere.

Concerning the issue of energies radiated off due to the interaction of black holes, as we showed here with a toy model, the changes imprinted by the Einstein-Born-Infeld black holes w.r.t. their classical counterparts may be significant, depending on $\alpha$ for a range of values of the fundamental parameter $b|Q|$. This could be important for gravitational wave detectors calibrated based on classical results. Besides, if it is possible to identify sources of radiation, then measurements upon such a quantity could give us information about electromagnetic interactions. We analyzed the radiated energies due to charged black hole interactions. This means that also electromagnetic radiation is always present in such processes. Identifying and analyzing this part of the radiation would gives direct information about astrophysical electrodynamical processes.

We further point out that all the above conclusions remain valid even in the case the systems present a slow rotation (when the rotational parameter $a\doteq J/M$, $J$ being the total angular momentum of the system as seen by distant observers, is much smaller than the outer horizon area or the mass of the hole). This is the case since the energy decomposition must be an even power of $a$, due to invariance requirements. Thereby, the previous analyses are in a sense stable against rotational perturbations.

Summing up, in this work we tried to emphasize the need of also taking into account the mass decomposition of a charged black hole for talking about the physical aspects it could display. Conceptually speaking this is of relevance since it could give us acumen of where and how to search experimentally for charged black holes and their interactions. In this regard, it would be also of interest to investigate the aspects of the electromagnetic radiation coming from the coalescence of charged black holes, because it could be much more easily observed, it would give us direct information about electromagnetic phenomena and of the coalescence process itself. It also seems that QPOs could also shed a light on the illation of black hole charges and the role played by the nonlinearities of the electromagnetism in the astrophysical scope, since they talk about phenomena that take place in the innermost regions of black holes (see \cite{1996csnp.book.....G} and references therein). We let this issue to be elaborated elsewhere.
%eliminated For simulations assessing the germaneness of nonlinear charged black holes into gravitational wave spectra, it would be of interest to scrutinize the energy decomposition for axially symmetric spacetimes, though this a toil.

\acknowledgments
We are indebted to Dr. Andrea Geralico for insightful comments and discussions within the theme of this work. J.P.P. acknowledges the support given by the Erasmus Mundus Joint Doctorate Program within the IRAP PhD, under the Grant Number 2011--1640 from EACEA
of the European Commission. J.A.R. acknowledge the support by the International Cooperation Program CAPES-ICRANet financed by CAPES -- Brazilian Federal
Agency for Support and Evaluation of Graduate Education within the Ministry of Education of Brazil.

%\bibliographystyle{apsrev}
%\bibliography{clothed_singularities}

\end{document}